\documentclass[aps,prl,floatfix,twocolumn,notitlepage,superscriptaddress,10pt]{revtex4-2}

\usepackage[table]{xcolor}
\usepackage{grffile}
\usepackage{amsmath,amsthm,amssymb,bbold}
\usepackage{bm}
\usepackage{microtype}
\usepackage{graphicx}   
\usepackage{verbatim}   
\usepackage{natbib}
\usepackage{enumitem} 
\usepackage{dsfont} 
\usepackage{hyperref}
\hypersetup{colorlinks,linkcolor=blue,urlcolor=blue,citecolor=blue}

\newcommand{\Tr}{\mathop{\rm Tr}}
\newcommand{\avg}[1]{\left\langle #1\right\rangle}

\newcommand{\subc}{\mathrm{C}}    
\newcommand{\subs}{\mathrm{S^z}}  

\begin{document}

\title{Spin--Charge Subordination in the Infinite-$U$ $SU(N)$ Hubbard Chain}

\author{C\u at\u alin Pa\c scu Moca}
\email{mocap@uoradea.ro}
\affiliation{Department of Physics, University of Oradea, 410087 Oradea, Romania}
\affiliation{Department of Theoretical Physics, Institute of Physics, Budapest University of Technology and Economics, H-1111 Budapest, Hungary}
\affiliation{MTA-BME Lend\"ulet ``Momentum'' Open Quantum Systems Research Group, Budapest University of Technology and Economics, H-1111 Budapest, Hungary}
\author{Ovidiu I. P\^{a}\c{t}u}
\affiliation{Institute for Space Sciences, Bucharest-M\u{a}gurele, R 077125, Romania}
\author{Gergely Zar\'and}
\affiliation{Department of Theoretical Physics, Institute of Physics, Budapest University of Technology and Economics, H-1111 Budapest, Hungary}
\author{Bal\'azs D\'ora}
\affiliation{Department of Theoretical Physics, Institute of Physics, Budapest University of Technology and Economics, H-1111 Budapest, Hungary}
\affiliation{MTA-BME Lend\"ulet ``Momentum'' Open Quantum Systems Research Group, Budapest University of Technology and Economics, H-1111 Budapest, Hungary}

\begin{abstract}
We obtain the exact full counting statistics of charge and flavor transport in the one-dimensional infinite-$U$ $SU(N)$ Hubbard model. 
The no-passing constraint freezes the ordered flavor sequence while the particle coordinates evolve as free spinless fermions. 
Consequently, charge transfer is governed by a free-fermion determinant,  whereas flavor 
transfer is exactly subordinated to the number of particles crossing the observation cut. For an arbitrary traceless Cartan generator 
this yields an exact relation between the charge and flavor cumulants. 
The complete asymptotic flavor distribution is a universal non-Gaussian M-Wright law. 
Together, these results establish spin--charge subordination as a kinematic mechanism for anomalously slow internal-state transport in
one-dimensional impenetrable quantum gases.
\end{abstract}

\maketitle


\paragraph{Introduction.\textendash}

The $SU(N)$ Fermi--Hubbard chain provides a paradigmatic setting for strongly correlated charge and flavor dynamics and is directly 
relevant to multicomponent ultracold-atom experiments~\cite{Hubbard1963,Gutzwiller1963,Kanamori1963,LiebWu1968,Sutherland1975,Schlottmann1994,Essler2005,Guan2013,Arovas2022,Cazalilla2009,Taie2012,Pagano2014,Scazza2014,Zhang2014,Hofrichter2016}. 
In the 
infinite-repulsion limit, double occupancy is forbidden and particles cannot pass one another in one dimension~\cite{Klein1973,Kumar2009,Mishra2001,KISHORE2004277}. Their spatial 
coordinates therefore evolve as spinless fermions while the ordered sequence of $SU(N)$ flavor labels remains frozen~\cite{OgataShiba1990,Essler2005,IPA98,GHPZ23,GQBZ24}.
The spin transferred during a given time interval is thus
determined by two distinct processes — the number of particles crossing the cut and the internal labels carried by those particles.



\begin{figure}[t]
 \includegraphics[width=\columnwidth]{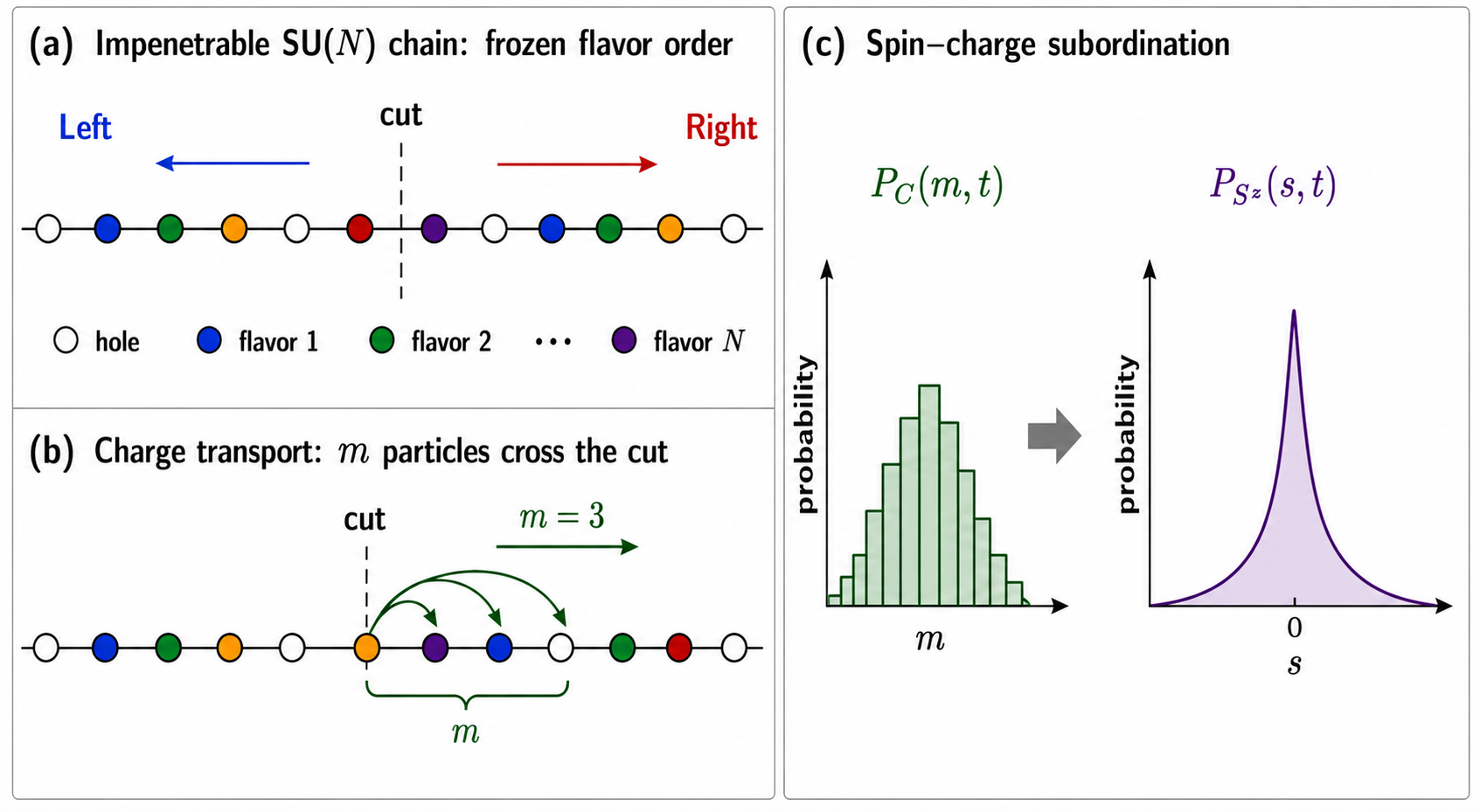}
 \caption{ Schematic of the spin--charge subordination in the infinite-$U$ (impenetrable) $SU(N)$ Hubbard chain. 
(a) The ordered sequence of $SU(N)$ flavor labels is frozen, while the particle coordinates evolve as free spinless fermions. 
(b) The number of particles crossing the central bond determines the transferred charge,
while the transported spin is the sum of the Cartan weights carried by those particles. (c) The spin generating function is 
subordinated to the charge generating function, as expressed in Eq.~\eqref{eq:spin-composition-preview}.
 }
\label{fig:sketch}
\end{figure}

Full counting statistics (FCS) probes this constrained dynamics beyond average currents by resolving the complete probability 
distribution of transferred conserved quantities~\cite{LevitovLesovik1993,Levitov1996,NazarovBlanter2009,KlichLevitov2009,Song2010,Song2012,Myers2020}. 
Recent work has revealed anomalous, strongly non-Gaussian current fluctuations in constrained systems, including 
scaling functions related to the M-Wright distribution~\cite{KSPI22,KVZ22,GMVK24a,GMV24,KSPP24a,KSIP24,KIPH25,YK25,YK26,YKBI26,UVGN26,pozsgay2026}. 
Exact microscopic generating functions in interacting quantum systems, however, remain rare. Earlier works by Krajnik and collaborators
established anomalous current-fluctuation statistics, single-file scaling laws, and M-Wright-type distributions in constrained and
integrable settings~\cite{KSPI22,KSPP24a,YK25,YK26}, while Ref.~\cite{Fujimoto2026} solved the corresponding problem in the spin-$1/2$
infinite-$U$ Hubbard chain using a generalized hydrodynamics approach. This naturally raises the question of whether the anomalous spin statistics are specific to $SU(2)$ or instead
reflect a symmetry-independent kinematic principle.

We show that the mechanism is in fact universal within the impenetrable $SU(N)$ family. Schematically, it is summarized in
Fig.~\ref{fig:sketch}: for arbitrary $SU(N)$ and any traceless Cartan generator, frozen flavor order yields an exact
\emph{spin--charge subordination}, namely that if a net number $m$ of particles crosses a cut, the transported flavor is the sum of the
Cartan weights carried by the corresponding $|m|$ consecutive particles in the frozen sequence. This gives the exact composition law

\begin{equation}
\chi_{\subs}(\lambda,t)=\sum_{m\in\mathbb Z}P_{\subc}(m,t)\,[g_N(\operatorname{sign}(m)\lambda)]^{|m|},
\label{eq:spin-composition-preview}
\end{equation}
where $g_N$ is the single-flavor characteristic function. Combining this relation with the exact free-fermion determinant for charge
FCS yields $\kappa_2^{\subc}\sim t$, $\kappa_2^{\subs}\sim t^{1/2}$, and a spin-transfer width $t^{1/4}$, together with a universal
non-Gaussian M-Wright scaling form. Compared with Ref.~\cite{Fujimoto2026}, our contribution is threefold: we extend the result from
$SU(2)$ to arbitrary $SU(N)$, provide a simpler analytical derivation of the anomalous spin fluctuations, and support the theory with
explicit MPS calculations for $SU(2)$, $SU(3)$, and $SU(4)$, giving an independent numerical verification of the predicted scaling~\cite{Vidal2007,SCHOLLWOCK201196,Valli2025,itensor}.
\paragraph{Model Hamiltonian.\textendash}
    
We consider the one-dimensional $SU(N)$ Hubbard model in the limit of infinite on-site repulsion, where double occupancy is 
completely forbidden. The Hamiltonian reads
\begin{equation}
H=-J\sum_{j=1}^{L-1}\sum_{\alpha=1}^{N}
\mathcal P
\left(
c_{j,\alpha}^{\dagger}c_{j+1,\alpha}
+\mathrm{H.c.}
\right)
\mathcal P ,
\label{eq:H}
\end{equation}
where $c_{j,\alpha}^{\dagger}$ creates a fermion of flavor $\alpha=1,\ldots,N$ at site $j$, $J$ is the nearest-neighbor hopping 
amplitude, and $\mathcal P =\prod_{i}\mathcal P_i$, with  $\mathcal P_i=\prod_{1\leq \alpha<\beta\leq N}\left(1-n_{i,\alpha}
n_{i,\beta}\right)$,  projects onto the physical Hilbert space without double occupancy. The local Hilbert space therefore 
contains only the empty state and the $N$ singly occupied states,
\begin{equation}
\mathcal H_j=
\left\{
|0\rangle,
|1\rangle,
\ldots,
|N\rangle
\right\},
\end{equation}
reducing its dimension from $2^N$ to $N+1$. 
Throughout this work we consider the infinite-temperature density matrix in the projected Hilbert space,
\begin{equation}
\rho_\infty=\frac{\mathcal P}{(N+1)^L},
\end{equation}
for which each allowed local configuration occurs with equal probability and the average particle density is $\bar n=N/(N+1)$.

The defining property of the model is that particles cannot pass through one another. Consequently, while the occupied coordinates 
propagate as free spinless fermions, the ordered sequence of $SU(N)$ flavor labels remains exactly conserved throughout the 
dynamics~\cite{Sutherland1975,OgataShiba1990,Schlottmann1994,Essler2005,Guan2013}. The many-body Hilbert space therefore factorizes 
into a charge sector, describing the motion of the particle coordinates, and a flavor sector consisting of a frozen string of flavors 
attached to these coordinates. This exact factorization is the microscopic origin of the spin--charge subordination established 
below: the charge dynamics is identical to that of free fermions, whereas the color dynamics is generated entirely by the random 
sequence of particles crossing the observation cut.

We investigate the integrated charge and flavor currents crossing the central bond of the infinite lattice using the 
two-projective-measurement protocol. For definiteness, we focus on the left subsystem, $L=\{\cdots,-1,0\},$ whose total 
particle number is
\begin{equation}
N_L=\sum_{j\le 0} n_j,
\qquad
n_j=\sum_{\alpha=1}^{N}
c_{j,\alpha}^{\dagger}c_{j,\alpha}.
\end{equation}
The transported flavor is associated with an arbitrary traceless Cartan generator of $SU(N)$ \cite{Georgi1999},
\begin{equation}
\Lambda_z=\mathrm{diag}\,(q_1,\ldots,q_N),
\qquad
\sum_{\alpha=1}^{N} q_\alpha =0,
\end{equation}
and the corresponding conserved quantity in the left subsystem is
\begin{equation}
S_L^{z}
=
\sum_{j\le 0}\sum_{\alpha=1}^{N}
q_\alpha n_{j,\alpha}.
\end{equation}
In the following, we derive the generating functions and the full counting statistics associated with these observables. Our 
approach exploits the exact conservation of the flavor ordering during the dynamics, which allows the spin generating function to 
be expressed analytically in terms of the charge generating function. In the impenetrable limit, the latter coincides with that 
of free fermions, yielding an exact solution for the joint charge and spin transport statistics.


\paragraph{Quantum generating function.\textendash}

To characterize charge and flavor transport beyond average currents, we use the quantum generating function (QGF), which encodes 
the complete probability distribution of a transferred conserved quantity and all of its cumulants ~\cite{LevitovLesovik1993,
Levitov1996,NazarovBlanter2009,KlichLevitov2009,CalabreseCardy2005,EislerPeschel2007,CalabreseEsslerFagotti2011,
CalabreseEssler2012,Stephan2013,Valli2025}. For either the left-half charge or $S^z$, $Q\in\{N_{L},S_{L}^{z}\}$,  we define the 
net transfer across the central bond during the interval $[0,t]$ as
$\Gamma_Q(t)=Q(t)-Q$. The corresponding two-time generating function is
\begin{equation}
\chi_Q(\lambda,t)
=
\Tr\!\left[
\rho_\infty e^{-i\lambda Q}e^{i\lambda Q(t)}
\right],
\label{eq:qgf}
\end{equation}
where $\lambda$ is the counting field. Because the infinite-temperature state commutes with $Q$, Eq.~\eqref{eq:qgf} coincides with 
the characteristic function obtained from the standard two-projective-measurement protocol. The full counting statistics follows 
by Fourier transformation,
\begin{equation}
P_Q(q,t)=
\int\frac{d\lambda}{2\pi}\,
e^{-i\lambda q}\chi_Q(\lambda,t),
\label{eq:FCS}
\end{equation}
with the integration interval chosen according to the periodicity of the corresponding spectrum.

The moments and cumulants are generated by derivatives at vanishing counting field,
\begin{align}
\mu_n^{(Q)}(t)
&=(-i)^n\left.\partial_\lambda^n\chi_Q(\lambda,t)\right|_{\lambda=0},\\
\kappa_n^{(Q)}(t)
&=(-i)^n\left.\partial_\lambda^n\ln\chi_Q(\lambda,t)\right|_{\lambda=0}.
\label{eq:moments-cumulants}
\end{align}
At the flavor-symmetric infinite-temperature point, the transfer distributions are symmetric, $P_Q(q,t)=P_Q(-q,t)$, and all odd 
moments vanish. The leading measure of spreading is therefore the second moment,
\begin{equation}
\mu_2^{(Q)}(t)
=
\avg{\Gamma_Q^2(t)}
=
\avg{[Q(t)-Q]^2},
\label{eq:secondmoment}
\end{equation}
which coincides with the second cumulant, $\kappa_2^{(Q)}(t)=\mu_2^{(Q)}(t)$, since $\avg{\Gamma_Q(t)}=0$. 
The exact generating functions obtained in the 
following sections determine not only these second moments but the complete charge and spin FCS.


\paragraph{Full counting statistics for charge.\textendash}

We first consider the QGF of the transferred charge, $\Gamma_{\subc}(t)=N_L(t)-N_L.$ In the impenetrable limit, the energy of an 
arbitrary eigenstate is independent of its spin configuration. Consequently, tracing over the spin degrees of freedom yields a 
Gaussian ensemble of free spinless fermions with filling $\bar n$ and diagonal correlation matrix 
$C_{ij}=\bar n\,\delta_{ij}$.
Let $ P_L=\sum_{j\le0}|j\rangle\langle j|$ denote the one-body projector onto the left half of the chain, and let
$ P_L(t)=u^\dagger(t)P_Lu(t) $
be its time evolution under the single-particle hopping Hamiltonian, where $ u_{jm}(t)=i^{m-j}J_{m-j}(2Jt),$ with $J_m(x)$ the 
Bessel function of the first kind \cite{AS64}. Employing Klich's trace formula~\cite{Klich2003,Schonhammer2007,AbanovIvanov2008}, 
the charge generating function~\eqref{eq:qgf} can be written as
\begin{equation}
\chi_{\subc}(\lambda,t)=
\det\!\left[ (1-\bar n)\mathbb{1} +\bar n\,e^{-i\lambda P_L}e^{i\lambda P_L(t)} \right].
\label{eq:charge-det}
\end{equation}
The determinant in Eq.~\eqref{eq:charge-det} involves the product of two orthogonal projections. Introducing the positive operator
\begin{equation}
X(t)=P_L\bigl[\mathbb{1}-P_L(t)\bigr]P_L,
\label{eq:charge-transfer-operator}
\end{equation}
which satisfies $ 0\le X(t)\le P_L\le\mathbb{1} $ its nonzero eigenvalues $p_a(t)\in[0,1]$ are precisely the single-particle 
transfer probabilities. An application of Halmos' two-projections theorem~\cite{Halm69,BS10} then yields the factorized 
representation
\begin{equation}
\chi_{\subc}(\lambda,t)=
\prod_a
\left[ 1-4 \bar n(1-\bar n)\sin^2\!\left(\frac{\lambda}{2}\right)p_a(t) \right].
\label{eq:charge-product}
\end{equation}
\begin{figure}[t]
 \includegraphics[width=0.8\columnwidth]{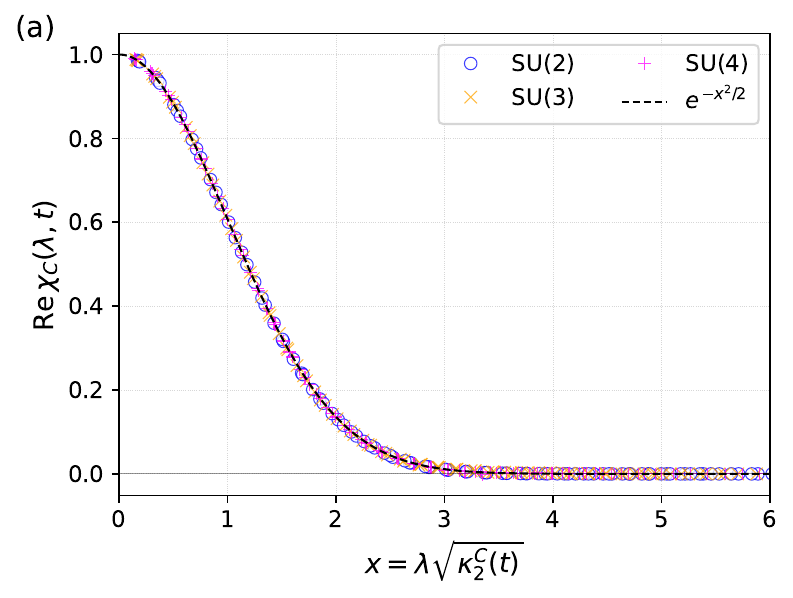}
 \includegraphics[width=0.8\columnwidth]{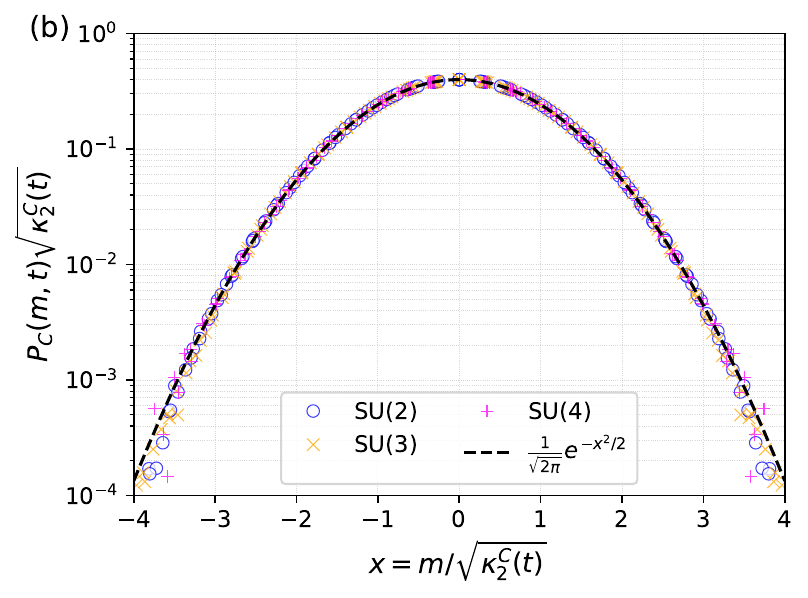}
 \caption{ (a) The generating function $\chi_{\subc}(\lambda,t)$ as function of the 
 rescaled counting field $\lambda\sqrt{\kappa_2^{\subc}(t)}$ for $SU(N=2,3,4)$. 
The dashed line is the Gaussian central-limit prediction, according to Eq.~(\ref{eq:sun_charge_fcs_gaussian}). 
 (b) The transferred-charge distribution $P_{\subc}(m,t)$ as function of the rescaled 
 charge transfer $m/\sqrt{\kappa_2^{\subc}(t)}$. The collapse follows the ballistic charge variance. 
 The dashed curve is the Gaussian central-limit prediction. Data is displayed on a logarithmic scale to highlight the Gaussian shape.
 Simulations use $L=200$ and bond dimension $M=128$.}
\label{fig:charge}
\end{figure}
Equations~\eqref{eq:charge-det} and~\eqref{eq:charge-product} are exact for arbitrary times. Moreover, since $\chi_{\subc}(\lambda,t)$ 
is an even function of $\lambda$, all odd charge cumulants vanish. Expanding Eq.~\eqref{eq:charge-product} to second order in $\lambda$ 
yields the variance of the transferred charge
\begin{equation}
\kappa_2^{\subc}(t)
=
2\bar n(1-\bar n)\,\Tr X(t).
\label{eq:charge-second-cumulant}
\end{equation}
Using the cyclicity of the trace together with the unitarity of the propagator, $\sum_j |u_{ij}(t)|^2=1,$
one finds $\Tr X(t)=\Tr P_L-\Tr\!\left[P_L(t)P_L\right]=\sum_{i\in L}\sum_{j\in R}|u_{ij}(t)|^2$ where $R=\{1,2,\ldots\}$ denotes the 
right half of the chain. Since  $|u_{ij}(t)|^2=J_{j-i}^2(2Jt)$  it follows that $\Tr X(t)=\sum_{r=1}^{\infty} r\,J_r^2(2Jt).$
Employing the identity
$
\sum_{r=1}^{\infty} r\,J_r^2(x)
=
\frac{x^2}{2}\left[J_0^2(x)+J_1^2(x)\right]
-\frac{x}{2}J_0(x)J_1(x),
$
together with the large-$x$ asymptotic expansion
$J_r(x)\sim \sqrt{\frac{2}{\pi x}} \cos\!\left(x-\frac{r\pi}{2}-\frac{\pi}{4}\right),$
we obtain the long-time behavior

\begin{equation}
\kappa_2^{\subc}(t)
\simeq
\frac{4}{\pi}\frac{N}{(N+1)^2}Jt
=
\hat{\kappa}_2^{\subc}Jt,
\qquad
Jt\gg1,
\label{eq:charge-second-cumulant-asymptotic}
\end{equation}
where $\hat{\kappa}_2^{\subc}=\frac{4}{\pi}\frac{N}{(N+1)^2}$ is the effective charge variance.

At long times the integrated current across the interface obeys a central-limit
form. In the regime $Jt\gg1$, the central part of the charge-transfer distribution is
therefore Gaussian, and the characteristic function reduces to
\begin{equation}
    \chi_{\subc}(\lambda,t)\simeq
    \exp\!\left[-\frac{\lambda^2}{2}\kappa_2^{\subc}(t)\right],
    \qquad |\lambda|\ll1,
    \label{eq:sun_charge_fcs_gaussian}
\end{equation}
The complete charge-transfer distribution follows from Fourier inversion, i.e. Eq.~\eqref{eq:FCS}, 
so at long times it approaches
\begin{equation}
P_{\subc}(m,t)
\simeq
\frac{1}{\sqrt{2\pi\kappa_2^{\subc}(t)}}
\exp\!\left[-\frac{m^2}{2\kappa_2^{\subc}(t)}\right].
\label{eq:charge-fcs-gaussian}
\end{equation}
Thus the charge-current variance is linear in time, while the rms transferred charge grows as $t^{1/2}$, with
an $N$-dependent prefactor fixed entirely by the constrained infinite-temperature
filling $\bar n$. 
In Fig.~\ref{fig:charge} we show the charge generating function and the corresponding transfer distribution 
for $SU(2)$, $SU(3)$, and $SU(4)$, together with the analytical estimates of Eqs.~\eqref{eq:sun_charge_fcs_gaussian} and 
\eqref{eq:charge-fcs-gaussian}, demonstrating the predicted scaling collapse and the approach to the Gaussian central-limit 
form.


\paragraph{Full counting statistics for spin.\textendash}

The crucial observation enabling an exact evaluation of the flavor (Cartan) QGF is that the ordered flavor sequence is frozen throughout the dynamics. As a consequence, spin transport is a stochastic process subordinated to charge 
transfer. If $|m|$ particles cross the central bond, the transported spin is the sum of $|m|$ independent Cartan weights. For a 
general traceless Cartan generator,  $ \Lambda_z=\mathrm{diag}(q_1,\ldots,q_N)$, each transferred particle carries Cartan weight 
$q\in\{q_1,\ldots,q_N\}$ with equal  probability $1/N$. The corresponding single-particle characteristic function is therefore
\begin{equation}
g_N(\lambda)
=
\langle e^{i\lambda q}\rangle
=
\frac{1}{N}\sum_{\alpha=1}^{N}e^{i\lambda q_\alpha}.
\label{eq:single-spin-characteristic}
\end{equation}

For fixed $m$, the flavor characteristic function is
$[g_N(\operatorname{sign}(m)\lambda)]^{|m|}$, where $m$ is positive (negative) if
particles enter (leave) the left subsystem.
Averaging over the charge-transfer distribution $P_{\subc}(m,t)$ gives the exact composition law already stated in Eq.~\eqref{eq:spin-composition-preview}.
Equivalently, the probability that a net spin $s$ is transferred across the central bond is
\begin{equation}
P_{\subs}(s,t)
=
\sum_{m\in\mathbb Z}
P_{\subc}(m,t)
\int_{-\pi}^{\pi}\frac{d\lambda}{2\pi}
e^{-i\lambda s}
\left[g_N(\lambda)\right]^{|m|}.
\label{eq:spin-prob}
\end{equation}
Equations~\eqref{eq:spin-composition-preview} and~\eqref{eq:spin-prob} constitute one of the main results of this work. They 
establish an exact mapping between the charge and spin full counting statistics, valid at arbitrary times for the balanced projected infinite-temperature ensemble considered here. The kinematic structure itself extends more generally whenever the frozen flavor sequence is sampled from the balanced sector.

The subordination mechanism also provides a simple derivation of the second cumulant of the transferred spin. Since the 
spin carried by  a single transferred particle has variance
\begin{equation}
\sigma_{\subs}^2
=
\frac{1}{N}\sum_{\alpha=1}^{N}q_\alpha^2
=
\frac{1}{N}\Tr\!\left(\Lambda_z^2\right),
\label{eq:spin-variance}
\end{equation}
the spin variance conditioned on the transfer of $|m|$ particles is simply $|m|\,\sigma_{\subs}^2$. For the standard 
normalization $\Tr(\Lambda_z^2)=1/2$, one has $\sigma_{\subs}^2=1/(2N)$. Averaging over the charge-transfer distribution 
then yields
\begin{equation}
\kappa_2^{\subs}(t)
=
\sigma_{\subs}^2
\sum_{m\in\mathbb Z}
|m|\,P_{\subc}(m,t)
=
\sigma_{\subs}^2
\avg{|\Gamma_{\subc}(t)|}.
\label{eq:spin-second-cumulant}
\end{equation}

For $Jt\gg1$ substituting the Gaussian central-limit form of $P_{\subc}(m,t)$ yields $\avg{|\Gamma_{\subc}(t)|}
\simeq\sqrt{2\kappa_2^{\subc}(t)/\pi}$, resulting in the asymptotic spin variance

\begin{equation}
\kappa_2^{\subs}(t)\simeq \sigma_{\subs}^2\sqrt{\frac{2 \kappa_2^{\subc}(t)}{\pi}} \simeq \sigma_{\subs}^2\frac{2}
{\pi}\frac{\sqrt{2N}}{N+1}\sqrt{Jt}.
\label{eq:spin-second-cumulant-asymptotic}
\end{equation}
Thus the integrated charge-current fluctuations obey $\kappa_2^{\subc}(t)\propto Jt$, whereas the subordinated spin-current 
fluctuations are parametrically slower, $\kappa_2^{\subs}(t)\propto (Jt)^{1/2}$. Equivalently, the rms spin transfer grows 
only as $(Jt)^{1/4}$~\cite{Bertini2016,CastroAlvaredo2016,Doyon2017,IlievskiDeNardis2017,Prosen2011,Ljubotina2017,Ljubotina2019,
GopalakrishnanVasseur2019,Spohn2014,Popkov2015,Krajnik2022}.

\begin{figure}[t]
 \includegraphics[width=0.8\columnwidth]{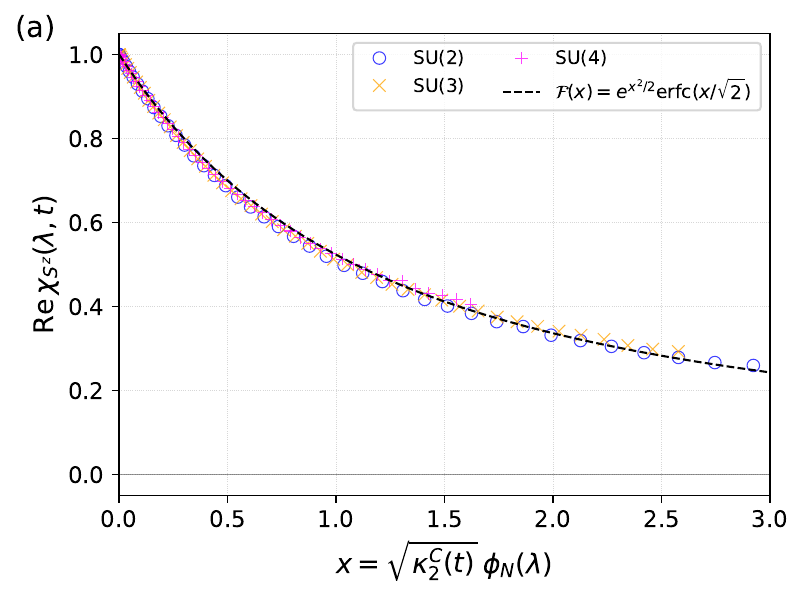}
 \includegraphics[width=0.8\columnwidth]{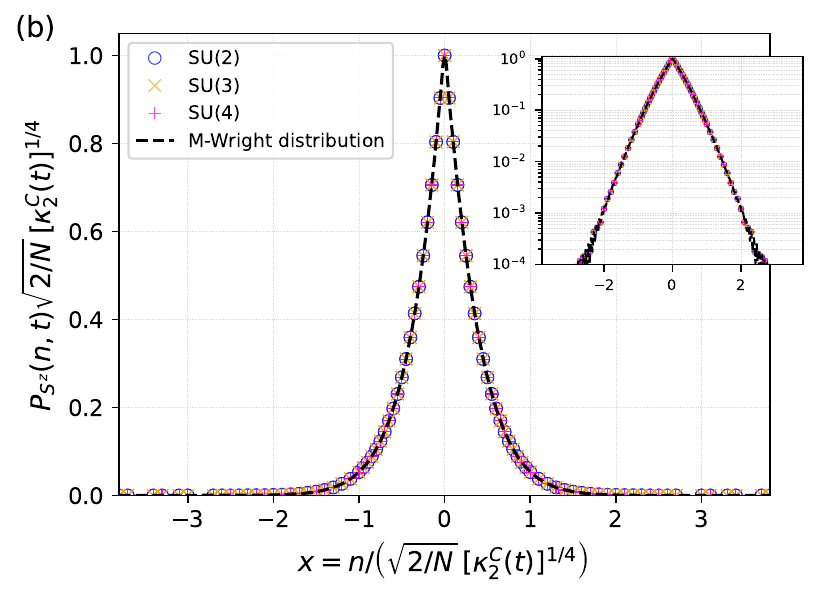}
 \caption{Spin full counting statistics across the central cut. (a) Spin generating function for $SU(2)$, $SU(3)$, and $SU(4)$, 
 rescaled according to Eq.~\eqref{eq:spin-scaling-form}. The dashed line denotes the analytical subordination scaling function 
 $\mathcal F(x)$. (b) Corresponding spin-transfer distributions plotted in the scaling variable set by 
 $\sqrt{\kappa_2^{\subs}(t)}$. The simulations use $L=200$ and bond dimension $M=128$. Data is displayed on a logarithmic scale 
 to highlight the non-Gaussian shape. The dashed line is the M-Wright scaling function of Eq.~\eqref{eq:mwright-distribution}.
 The inset represents the same data on a logarithmic scale. The time values are fixed to $Jt=\{90,100\}$.
}
 \label{fig:spin}
\end{figure}


In the scaling limit $Jt\to\infty$ with $\sqrt{\kappa_2^{\subc}(t)}\,\phi_N(\lambda)$ fixed, the discrete sum in 
Eq.~\eqref{eq:spin-composition-preview} becomes an integral, yielding
\begin{equation}
\chi_{\subs}(\lambda,t)
=
\mathcal{F}\!\left( \sqrt{\kappa_2^{\subc}(t)}\,\phi_N(\lambda)\right),
\label{eq:spin-scaling-form}
\end{equation}
where $\mathcal{F}(x)=e^{x^2/2}\,\mathrm{erfc}\!\left(\frac{x}{\sqrt{2}}\right)$ and  $\phi_N(\lambda)=-\ln g_N(\lambda).$

Expanding $\ln\chi_{\subs}(\lambda,t)$ around $\lambda=0$ reveals an infinite hierarchy of nonvanishing higher-order cumulants,
$\kappa_4^{\subs}\propto t, \kappa_6^{\subs}\propto t^{3/2},\dots,$ demonstrating that spin fluctuations are intrinsically 
non-Gaussian. Their physical origin lies in the compound nature of spin transport: charge fluctuations determine the stochastic 
operational time, while the frozen flavor ordering assigns independent random Cartan weights to the transferred particles.

The scaling form of the spin-transfer probability is obtained by Fourier transforming Eq.~\eqref{eq:spin-scaling-form} and 
introducing  the rescaled variables $\mathcal{J}_{\subc}=m/\sqrt{Jt}$, $\mathcal{J}_{\subs}=s/(Jt)^{1/4}.$
Using the small-$\lambda$ expansion $\phi_N(\lambda)=-\ln g_N(\lambda)\simeq \sigma_{\subs}^2\lambda^2/2,$ the Fourier integral 
can be evaluated analytically, with the result
\begin{equation}
P_{\subs}\!\left[(Jt)^{1/4}\mathcal{J}_{\subs},t\right]
\simeq
\frac{1}{(Jt)^{1/4}}\frac{1}{\sigma_{\subs}}
\,
\mathbb{P}_{\rm MW}\!\left[\frac{\mathcal{J}_{\subs}}{\sigma_{\subs}},\sigma_{\subc}\right],
\label{eq:spin-mwright-scaling}
\end{equation}
where $\sigma_{\subc}^2=\hat{\kappa}_2^{\subc}$ [see Eq.~\eqref{eq:charge-second-cumulant-asymptotic}], and the M-Wright 
distribution~\cite{Wright1933,Mainardi2010,Krajnik2022} is defined by
\begin{equation}
\mathbb{P}_{\rm MW}[\mathcal{J},\sigma]
=
\frac{1}{\pi\sigma}\int_0^\infty
\frac{d\mathcal{J}_{\subc}}{\sqrt{\mathcal{J}_{\subc}}}
\exp\!\left[ -\frac{\mathcal{J}_{\subc}^2}{2\sigma^2}-\frac{\mathcal{J}^2}{2\mathcal{J}_{\subc}} \right].
\label{eq:mwright-distribution}
\end{equation}
Equation~\eqref{eq:spin-mwright-scaling} demonstrates that the integrated spin current of the impenetrable $SU(N)$ Hubbard chain 
exhibits anomalous fluctuations governed by the strongly non-Gaussian M-Wright distribution. It thus extends the recently 
discovered $SU(2)$ result of Ref.~\cite{Fujimoto2026} to the entire $SU(N)$ family.

Figure~\ref{fig:spin} confirms both the analytical scaling of the generating function and the M-Wright collapse of the spin 
distribution for $SU(2)$, $SU(3)$, and $SU(4)$ matrix-product-state simulations.


\paragraph{Finite temperatures.\textendash}

The spin--charge factorization of the eigenstates is unaffected by thermal averaging. Consequently, the spin--charge subordination 
mechanism remains valid at finite temperature, and the exact relations~\eqref{eq:spin-composition-preview} and~\eqref{eq:spin-prob} 
continue to hold. At finite temperature, however, the correlation matrix $C_{ij}$ is no longer diagonal, and therefore the 
derivation of the asymptotic charge variance leading to Eq.~\eqref{eq:charge-second-cumulant-asymptotic} is no longer applicable. 
Instead, the charge variance can be obtained from the equilibrium Levitov--Lesovik formula~\cite{Schonhammer2007}. In the impenetrable 
limit, charge transport of balanced multicomponent systems is equivalent to that of free fermions with a renormalized chemical potential, 
$\mu'=\mu+T\ln N$. For perfect transmission, the two subsystems are characterized by the same Fermi distribution,
$ f(k)=\frac{N}{N+e^{-\beta(2J\cos k+\mu)}}$. 
One then finds
\begin{equation}
\hat\kappa_2^{\subc}(\beta)
=
\frac{N}{J\pi\beta}
\left(
\frac{1}{N+e^{-\beta(2J+\mu)}}
-
\frac{1}{N+e^{\beta(2J-\mu)}}
\right),
\label{eq:finite-temperature-charge-variance}
\end{equation}
where $\beta=1/T$ is the inverse temperature. Equation~\eqref{eq:finite-temperature-charge-variance} is valid at all finite 
temperatures, except in the vicinity of zero temperature, where the Levitov--Lesovik formula for perfect transmission ceases to 
be applicable.
Temperature affects only the nonuniversal prefactor $\hat\kappa_2^{\subc}$; the universal scaling forms of both the charge and 
spin full counting statistics remain unchanged (see also Ref.~\cite{Fujimoto2026}).


\paragraph{Early-time dynamics.\textendash}

Equation~\eqref{eq:spin-composition-preview} is an exact consequence of the spin-incoherent nature of the system and constitutes 
a microscopic identity of maximum generality. It is valid for all temperatures, including zero temperature provided that the limit 
$U\rightarrow\infty$ is taken before $T\rightarrow0$, at all times, and for any $SU(N)$ Cartan generator. This underscores the 
generality of our approach, in contrast to other methods available in the literature.
In the short-time single-hop regime $Jt\ll1$, only $m=0,\pm1$ contribute at leading order. Writing $p_t=\bar n(1-\bar n)(Jt)^2$, the 
charge and spin QGFs reduce to
\begin{align}
\chi_{\subc}(\lambda,t)&=1-2p_t(1-\cos\lambda)+\mathcal O((Jt)^4),\\
\chi_{\subs}(\lambda,t)&=1-p_t\left[2-g_N(\lambda)-g_N(-\lambda)\right]+\mathcal O((Jt)^4).\nonumber
\end{align}
Equivalently, the charge FCS is trinomial, while the spin FCS is a dilute symmetric mixture of the single-particle Cartan weights $\pm q_\alpha$. 
Thus both sectors are initiated by the same rare hopping events at short times, whereas the distinct large-time 
scaling of the two sectors emerges only after many transfers have accumulated.

\paragraph{Experimental prospects.\textendash}

The predicted charge and spin full counting statistics should be directly observable in multicomponent ultracold Fermi gases and
quantum-gas microscopes, since our results remain valid at finite temperature. Alkaline-earth-like systems such as $^{173}$Yb realize
$SU(N)$ Hubbard chains with tunable filling and interactions~\cite{Pagano2014,Scazza2014,Zhang2014,Hofrichter2016}, while spin-resolved
microscopy of $^{87}$Sr already provides single-atom, state-resolved readout across the full nuclear-spin manifold~\cite{BuobPRXQ,gasferrer}.
Repeated projective snapshots after evolution time $t$ would therefore reconstruct the charge and Cartan-spin transfer distributions directly,
revealing the contrast between ballistic charge broadening and the anomalous $t^{1/4}$ spin scale, as well as the collapse onto the universal
M-Wright form. Similar sampling protocols are also natural in programmable quantum simulators~\cite{alam,tarruell2018}, so the exact scaling
laws derived here provide quantitative benchmarks for both analog and digital platforms in the strongly interacting regime.

\paragraph{Conclusions.\textendash}
We have shown that anomalous flavor full counting statistics in the infinite-$U$ $SU(N)$ Hubbard chain follows from a simple kinematic 
principle. The no-passing constraint freezes the ordered flavor sequence, so charge transfer fixes the number of transported particles 
while their frozen Cartan weights determine the flavor transfer. This yields an exact subordination relation between the two generating 
functions. The free spinless charge sector has $\kappa_2^{\subc}\propto t$, but the additional random sum over the transported flavor 
labels produces $\kappa_2^{\subs}\propto t^{1/2}$, an rms spin transfer $\propto t^{1/4}$, and a universal non-Gaussian M-Wright 
distribution. The mechanism is independent of the particular value of $N$ and therefore identifies a general route by which 
impenetrability can generate anomalously slow internal-state transport. The predicted FCS can be tested directly in multicomponent 
ultracold atomic systems and quantum simulators~\cite{Kinoshita2006,RigolMuramatsuOlshanii2006,Rigol2007,VidmarRigol2016,GarrisonGrover2018}.

\begin{acknowledgments}

This work was supported by the National Research, Development and Innovation Office - NKFIH Project No. K142179, by a grant of 
the Ministry of Research, Innovation and Digitization, CNCS/CCCDI-UEFISCDI, under projects number PN-IV-P1-PCE-2023-0159 and 
PN-IV-P1-PCE-2023-0987. 
This work  was also supported by the HUN-REN Hungarian Research Network through the Supported Research Groups Programme, 
HUN-REN-BME-BCE Quantum Technology Research Group (TKCS-2024/34). O.I.P. acknowledges financial support from Grant No.
30N/2023, provided through the National Core Program of the Romanian Ministry of Research, Innovation, and Digitization.
We acknowledge the Digital Government Development and Project Management Ltd. for awarding us access to the Komondor HPC facility 
based in Hungary. We acknowledge the use of the computing infrastructure provided by the University of Oradea and by IOSIN-PACTES
at the Institute of Space Science - INFLPR Subsidiary, Bucharest-M\u{a}gurele, Romania.
G.Z. acknowledges financial support from the National Research, Development and Innovation Office - NKFIH Advanced Grant No. 152794

\end{acknowledgments}

\bibliographystyle{apsrev4-2}
\bibliography{references}

\end{document}